# Image-based study of granular column collapse over controlled-roughness surfaces

Étude sur l'effondrement d'une colonne granulaire sur une base de rugosité contrôlée le traitement d'images

**Shuocheng Yang**
*Department of Civil Engineering, The University of Hong Kong, Pokfulam, Hong Kong, yangsc93@hku.hk*

**Lu Jing**
*Department of Chemical and Biological Engineering, Northwestern University, Evanston, IL 60208, USA*

**Chung Yee Kwok**
*Department of Civil Engineering, The University of Hong Kong, Pokfulam, Hong Kong*

**Gengchao Yang**
*School of Aeronautics and Astronanutics, Sun Yat-Sen University, 510275, Guangzhou, China*

**Yuri Dumaresq Sobral**
*Departamento de Matemática, Universidade de Brasília, Campus Universitário Darcy Ribeiro, 70910-900, Brasília, Brazil*

ABSTRACT: Basal effects have important implications for the high mobility and long runout of granular flows such as rock avalanches and landslides. However, fundamental understanding of the basal effect in granular flows remains challenging due to the complex forms of base roughness and the multiscale nature of flow-bed interactions. Here we experimentally investigate the basal effect in granular column collapse over controlled-roughness bases. Image processing methods are developed to obtain robust measurements of base roughness, runout distance and deposit morphology. A geometric roughness parameter $R_a$ is applied to consider both the size and spatial distribution of base particles, which enables systematic analysis of the basal effect. The results indicate that the runout distance can be characterized as a function of $R_a$, regardless of the variations in the base particle size and spacing, and the roughness has a major influence on the frontal region of the granular flow, as well as the overall profile of the granular deposit. When $R_a$ is increased beyond a threshold value ($R_a > 0.62$), flow characteristics show minor changes, which coinsides with a previous phase diagram for the transition between slip and non-slip boundary conditions from steady state granular flow simulations.

RÉSUMÉ : L'effet de la surface a des implications importantes sur la mobilité élevée et la distance parcourue par des écoulements granulaires comme les avalanches rocheuses et les glissements de terre. Néanmoins, la compréhension fondamentale de l'effet de la surface dans les écoulements granulaires reste encore difficile à cause des formes complexes de sa rugosité et de la nature multi-échelle des interactions entre l'écoulement et la surface rugueuse. Dans ce travail, nous étudions expérimentalement l'effet de la rugosité de la surface dans des surfaces avec rugosité contrôlée. Des méthodes de traitement d'images ont été développées pour l'obtention des mesures robustes de la rugosité de la surface et des propriétés de l'écoulement. Le paramètre de rugosité géométrique Ra est proposé pour considérer à la fois la taille et la distribution des particules de la surface, ce qui permet une analyse systématique de l'effet de la rugosité sur l'écoulement granulaire. Les résultats indiquent une influence importante de Ra sur l'écoulement granulaire. Lorsque Ra > 0,62, les caractéristiques de l'écoulement ne sont presque plus affectées, en concordance avec le diagramme de phase pour la transition entre les conditions de glissement et de non-glissement dans les simulations d'écoulements granulaires en régime permanent.

## 1 INTRODUCTION

Rock avalanches, debris flows and other flow-like landslides involve granular materials flowing over rough bases, where the base roughness plays a crucial role in modifying the flow dynamics and impact power. For instance, in the study of rock avalanches, both in-situ data and simulation results show that the basal material controls the flow mobility over other mechanisms like source volume (Aron and McDougall, 2019); in small-scale debris flow experiments, base roughness is necessary for producing realistic deposit morphologies with levee formation (de Haas et al. 2015); and in flume tests, increasing basal friction results in a lower flow front velocity and a smaller dynamic impact force to rigid barrier (Ahmadipur et al. 2019).

The interaction between a granular flow and a rough base is a multiscale process, where the flowing particle diameter, flow dimensions, and the roughness size of the bumpy surface are all relevant length scales. It is well known that granular flows tend to slide along flat and less rough bases, but no-slip condition occurs when the base is sufficiently rough (Goujon et al. 2003, Jing et al. 2016). To systematically study the basal effect in granular flows, a common approach to producing (geometric) base roughness in laboratory tests is to glue a layer of particles on the boundary surface (Pouliquen 1999, Santomaso and Canu 2001, Goujon et al. 2003). By varying the size and spacing of the base particles, Goujon et al. (2003) constructed different geometric roughness for granular flows down inclined planes. Their tests revealed that, apart from the influence of the particle size, the compactness of base particles also controls the final deposit length, implying a complicated dependence of the flow dynamics on both the particle size and spatial arrangement. In discrete element method (DEM) simulations, the size and spatial distribution of base particles can be easily controlled to generate a wide range of geometric base roughness (Sibert et a. 2002, Weinhart at al. 2012, Kumaran and Maheshwari 2012, Jing et al 2016). In our recent study (Jing et al. 2016), we developed a dimensionless roughness parameter ($R_a$) that can describe the base roughness of fixed equally-size spheres with either regular or random packing in a unified manner. We found in DEM simulations of mono-disperse steady flows that, as $R_a$ increases, a transition from slip ($R_a < 0.51$) to non-slip ($R_a > 0.62$) conditions occurs at the base. The same phase transition characterized by $R_a$ was observed in bi-disperse flows where particle segregation occurs (Jing et al. 2017).

Previous studies of the basal effect of granular flows focused mainly on steady state problems. However, transient granular flows may exhibit different behaviors at the boundary, including unexpected slip over seemingly rough substrates that enhances the flow mobility (Domnik and Pudasaini, 2012). To study transient granular flow behaviors, a simplified model test of granular column collapse has been widely adopted (Lagrée 2011, Jing et al. 2018, Yang et al. 2019). However, no experimental study on granular column collapse has systematically controlled the base roughness and considered both size and spatial distribution of the particle-glued base, and the basal effect in granular column collapse remains poorly understood. In this study, we conducted column collapse tests over carefully controlled particle-glued bases. We pay special attention to image processing to obtain robust characterization of the compactness of the glued base particles, hence the base roughness $R_a$, and to achieve accurate measurement of the runout distance and the deposit profile of the granular flows. We analyze experimental results considering the roughness parameter $R_a$, which enables a unified description of the base roughness effect on the runout behavior of granular column collapses, and a phase transition is observed where the roughness effect tends to saturate as $R_a$ exceeds a threshold value.

The paper is organized as follows: In Sect. 2, we briefly introduce the characterization of base roughness for granular flows. Then, we present the experimental set-up and the detection algorithms used in image processing in Sect. 3. Section 4 shows the test results and conclusions are drawn in Sect. 5.

## 2 CHARACTERIZATION OF BASE ROUGHNESS

For a bumpy base made of mono-layer fixed particles, as adopted in this study, we assume that the centers of all particles are coplanar. Thus, the particles assembly can be discretized into triangular areas using the Delaunay triangulation scheme (in Fig. 1a). As shown in Fig.1b, the gray triangle represents a void area $A_i$ located right in the center area formed by three base particles (whose centroids are the vertices of the triangle), and $A_i$ can be expressed as (Jing et al. 2016):

$$A_i = d_b^2 \sqrt{\frac{\Sigma_k(1+\varepsilon_k)}{2} \prod_k [\frac{\Sigma_k(1+\varepsilon_k)}{2} - (1+\varepsilon_k)]} \quad (1)$$

where $d_b(1+\varepsilon_k)$ is the spacing, i.e., the length of each side of the triangle in Fig. 1b, $\varepsilon_k$ represents $l_k/d_b$ in Fig. 1b, with $k = 1, 2, 3$, and $d_b$ is the base particle diameter. Since base roughness can be interpreted as the resistance of the local void to the motion of flowing particles over this void, a maximum resistance (roughness) is assumed when a flowing particle can be exactly entrapped by the void. In this case (Fig. 1c), as the centroid of flowing particle is coplanar with the three base particles, the area of the equilateral triangle $A_m$ can be calculated as

$$A_m = d_b^2 \frac{3\sqrt{3}(1+\lambda)^2}{16} \quad (2)$$

where $\lambda = d/d_b$ is the flow-to-base particle size ratio and $d$ is the flowing particle diameter.

For an arbitrary mono-layer of fixed base particles, we define the (mean) roughness as:

$$R_a = \frac{1}{N}\sum_{i=1}^{N}\frac{A_i}{A_m} \quad (3)$$

where $N$ is the number of triangular voids. Thus, $R_a$ is a function of $\varepsilon_k$ and, with lower and upper limits of zero (no geometric roughness) and unity (maximum roughness), respectively. Note that when $A_i > A_m$, underneath layers (either a flat plane or a granular layer) can also contribute to the base roughness, as discussed in detail by Jing et al. (2016). However, this $A_i > A_m$ scenario is not considered in this study for simplicity.

In practice, the concept above can be implemented in a simpler way via compactness $c$, which is the ratio between the projected area occupied by all the base particles and the total area. As shown in Fig. 1c, $A_i$ in Fig. 1b can be projected as Fig.1d where the void area $A_i$ equals to the gray area plus the white area. The equivalent local void area (gray area plus white area) can be expressed in terms of local compactness $c_i$ and $d_b$ by considering 2D geometry in Fig. 1d:

$$A_i = \frac{\pi}{8c_i}d_b^2 \quad (4)$$

Assuming that the overall packing of the base follows the Delaunay triangulation scheme, we can extend Eq. 4 to the whole base as $A = d_b^2\pi/(8c)$. Therefore, the base roughness can be approximated as $R_a \approx A/A_m$, which is a function of the compactness $c$ and the size ratio $\lambda$. It is evident from this formulation that a looser packing or larger size of the base particles tends to increase the geometric roughness.

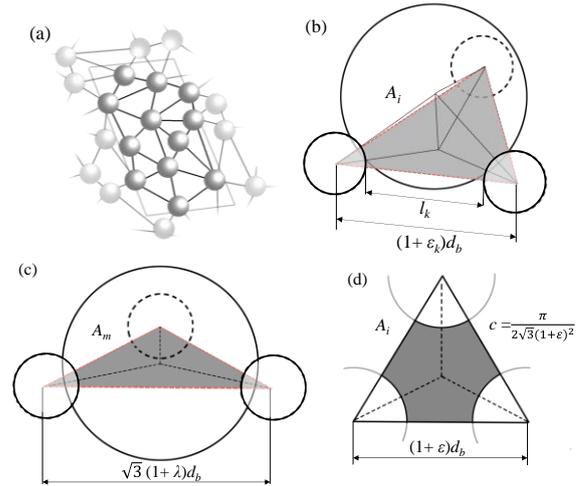

Figure 1. Definition of base roughness. The larger sphere represents a flowing particle, and the smaller sphere are the fixed base particles. (a) Discretization of a particle-glued base. (b) An arbitrary void area $A_i$. (c) An equilateral void area $A_m$, representing the scenario of the maximum roughness. (d) A projected equilateral void area $A_i$.

## 3 EXPERIMENTAL METHODS

### 3.1 *Experimental set-up and rough base preparation*

Figure 2 shows the experimental setup for the collapse of granular columns over controlled-roughness bases. The tank consists of Perspex planes with dimensions 50 cm, 30 cm, and 20 cm in the $x$, $y$, and $z$ directions, respectively. A vertically positioned 1.8 mm thick aluminum gate is constrained by two slots on the side walls (parallel to the $yz$-plane), which can be rapidly removed by releasing a dead weight. There are three pairs of slots built in the tank, but here we focus on the one at $x = 5$ cm, leading to a granular column with initial length $L_i = 5$ cm. The column body was prepared by pouring spherical glass beads with a mean diameter of 1.436 mm and density of 2468 kg/m³ into the container. The tank is then tapped till the column top surface was flat; in this work, the initial column height is fixed at $H_i$ = 5cm (i.e., an initial aspect ratio of 1). The total mass of the glass beads was controlled such that the initial packing density is 0.67 in all experiments.

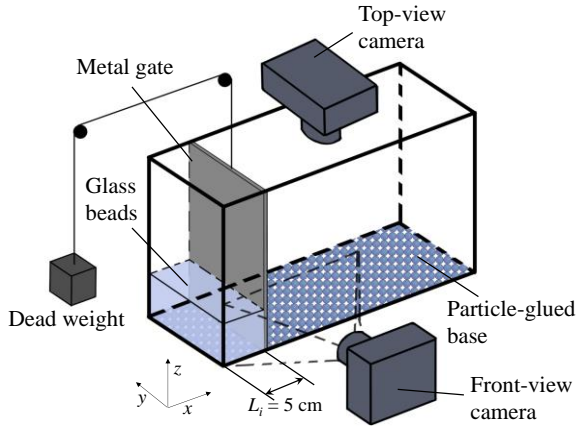

Figure 2. Sketch of the experimental set-up.

Base roughness was produced by gluing particles on a black plastic plane (of size 20 cm × 50 cm) with double-size tapes covered. First, base particles of a designated diameter $d_b$ were poured over the inclined sticky plane, where the slope angle is varied to produce different compactness of the base particles. After distributing the base particles evenly, a very thin layer of epoxy resins was carefully applied at the top of the base particles to prevent erosion during the test. Note that we adopted epoxy resins with a high liquidity to ensure that it sinks at the bottom with negligible damage to the basal surface. Once the resin dried out, in order to facilitate image processing, very thin black pigment was painted over the base particles. Examples of close-up binary photos of particle glued bases are shown in Fig. 3. Following the procedures above, we controlled the particle size and its spatial distribution on the glued-particle base (image-based characterization of the base roughness in light of $R_a$ is elaborated below).

After preparation of the granular column on a controlled-roughness base, the metal gate was lifted rapidly. The granular column collapsed onto the horizontal base and propagated in the *x*-direction. During each test, filming camera was set in front of the tank facing perpendicularly towards the *xz*-plane (Fig. 2), which has an image resolution of 3840 × 2160 pixels. An illumination light source was placed in the left front of the tank. After each test, the same camera was used to capture the top view of the deposition above the tank. For each type of the rough base, we conducted three repeated tests to obtain average results.

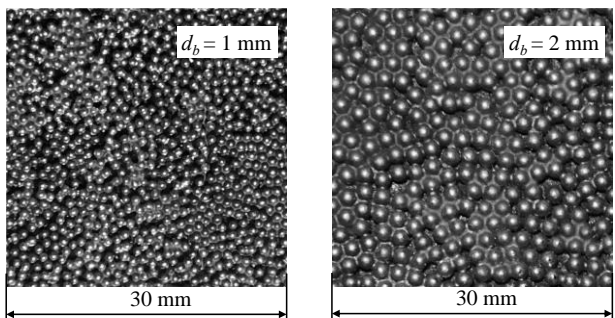

Figure 3. Photos of particle-glued bases.

### 3.2 Image processing

Image-processing algorithms are developed using the library of functions, OpenCV (Bradski and Kaehler 2008), with the open-source program ImageJ (Abràmoff et al. 2012).

#### 3.2.1 Determination of the base particle compactness

To determine the compactness as well as the spacing between base particles, it is necessary to locate the center of each base particle. A correlation algorithm between the template image and the target image was used (Gonzalez and Woods 1992). The photos of controlled-roughness bases were taken using the same camera as in Fig. 2. The processing steps are illustrated in Fig. 4:

(a) Four pairs of squared areas as the measurement cell with length of $10d_b$, $20d_b$, and $30d_b$ were selected along the longer side of the base. Results show that the variation of measured area or location has little influence on the value of compactness. Thus, all measurement cells were taken in the calculation of the mean compactness, its standard deviation and the mean spacing.

(b) After camera calibration, template image of a single glass bead was selected from the initial image (Fig. 4a). Note that the size of the template glass bead is slightly smaller than the actual size of glass bead to reduce object distortion due to reflection. As long as the central reflective point and surrounding dark area were captured, results will be independent on the selection of the template image.

(c) An algorithm to search the template was applied. Then, we obtained an image showing the zones of high and low correlations in dark and light gray, respectively (Fig. 4b). In Fig. 4b, an "dark cloud" can be seen for each bead in the initial image. The local maximum value was selected in each "dark cloud" and was regarded as the position of base particle. Then the particle positions are determined in Fig. 4c.

(d) The coordinates of these maximum values were extracted to generate spheres with the same size of base particles as shown in Fig. 4d. Because the calculation of the correlation coefficient is not accurate around the image edges, we eliminated the detection around the edges. With every particle projected in the 2D plane, compactness is calculated as the area occupied by particles (black area) divided by the total area (area outlined by the red box in Fig. 4d). As the particle positions were obtained, the mean spacing between particles were also calculated.

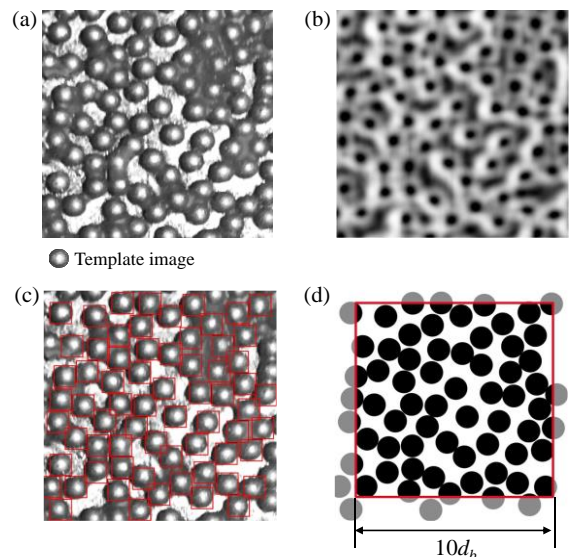

Figure 4. Detection of base particle positions ($d_b$ = 3 mm).

### 3.2.2 *Detection of the runout distance*

Runout distances were obtained from the top-view photo (Fig. 5a) of the deposition. An analysis of binary pixel intensity was applied. The processing steps were explained based on Fig. 5 as follow:

(a) After camera calibration, a rectangular area of 10 cm × 20 cm was selected (outlined by red solid line) in the center of the photo (see Fig. 5b).
(b) Initial image was set as binary and was thresholded. The threshold value was chosen from the trial image, in which one layer of particles at the base. The average gray level was extracted as the threshold grey level. This step is to make sure that the threshold represents the boarder of the continuum granular body. In our study, when the front of the flow is as shallow as one layer of particle, particle bouncing instead of flowing dominates the frontal region, which is filtered out.
(c) The measured area in Fig. 5b was cut into stripes as one of them as shown aside. The number of the stripes was set as 30, 45, 60, in order to keep the width of each stipes larger than 2 times of particle size. The pixel intensity (the area percentage of non-zero pixel) was extracted for each stripe.
(d) As the pixel intensity of each stripe is plotted in Fig. 5c, error function was used to fit the scattered extracted percentage. According to Fig. 5c, the number of stripes does not affect the fitting results. Then, the $x$-coordinate of mean value of error function was regarded as the inflection point as well as the runout distance (Fig. 5c). Meanwhile, the standard deviation of the runout measurement was obtained.

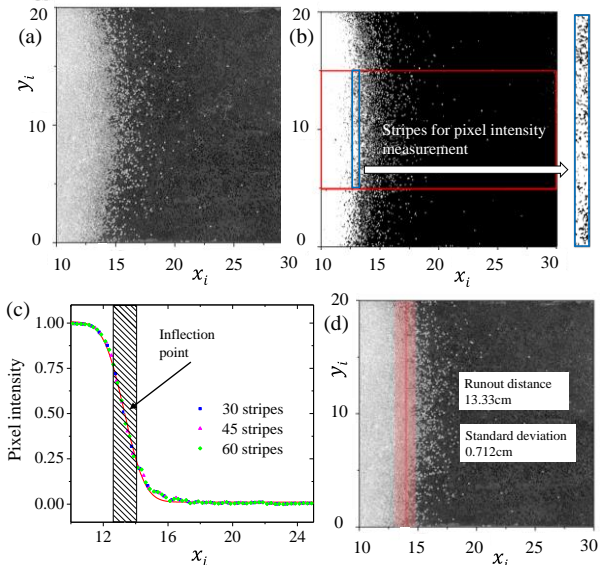

Figure 5. Detection of the runout distance.

### 3.2.3 *Detection of the deposit profile*

The depth profile of the final deposit was obtained from the front-view photo (Fig. 6a). Classic image-processing operators were used (Soille 1999). The processing was done in several steps as listed below:

(a) The front-view image was calibrated, and other trivial parts were eliminated.
(b) After the image was reset as binary, the contrast of image was adjusted by stretching the hologram of grey level using ImageJ (Fig. 6a). Then, the threshold value was set according to as the function of the luminosity and contrast of the source image (Soille 1999).
(c) Apart from continuous granular body, other white zones with small pixels were eroded for four iterations so the white dot on the top would disappear (Fig. 6b).
(d) By applying contouring algorithm, we obtained hundreds of scattered dots outlining the profile (Fig. 6c). All the scattered dots from three test for each case were extracted and used to interpolate the profile (Fig. 6d). The runout distance obtained from Sec. 2.2.2 was invoked to determine the front of the final profile.

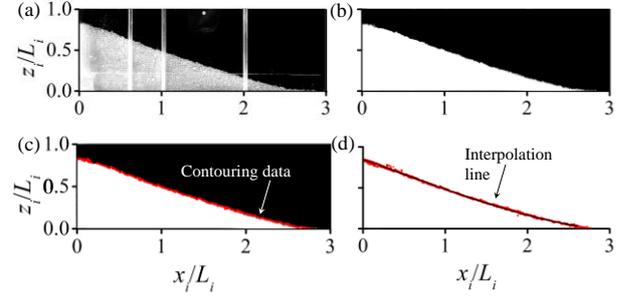

Figure 6. Detection of the final deposit profile.

## 4 RESULTS AND DISCUSSION

In this study, we constructed four types of the particle-glued base. By applying the image processing from Sec. 3.2.2, we obtained the compactness $c$ and mean spacing $\varepsilon$ for each base. We also obtained the spacing $\varepsilon_c$ deduced from $c$ with the relation shown in Fig. 1d, which agrees well with the measured spacing $\varepsilon$. As shown in Table 1, $R_a$ of these bases ranges from 0.29 to 0.89, spanning a wide range of roughness conditions (Jing et al. 2016). Among all five bases, Base 1 is the smoothest, which consists of the smallest base particle, while Base 5 with 3 mm base particle is the roughest. For Base 3 and 4 that have the same base particle size, a higher compactness (smaller spacing) of Base 3 results in a smaller $R_a$, indicating that Base 3 is smoother than Base 4. Note that Base 2 and Base 3 have very close $R_a$ even though their $d_b$ and $c$ are different. Note also that Base 2 is constructed using the same particle as the flowing particles (diameter 1.436 mm ± 10%).

Table 1. Parameters of the particle-glued base.

| Base ID | 1 | 2 | 3 | 4 | 5 |
|---|---|---|---|---|---|
| $d_b$ (mm) | 1 | 1.436 | 2 | 2 | 3 |
| $\varepsilon$ | 0.152 | 0.229 | 0.053 | 0.220 | 0.182 |
| $c$ (standard deviation) | 0.683 (0.028) | 0.602 (0.033) | 0.821 (0.023) | 0.612 (0.031) | 0.651 (0.032) |
| $\varepsilon_c$ | 0.151 | 0.227 | 0.051 | 0.217 | 0.180 |
| $R_a$ (standard deviation) | 0.298 (0.002) | 0.502 (0.002) | 0.499 (0.002) | 0.669 (0.003) | 0.850 (0.003) |

Figure 7 shows the final deposits of collapsed columns on the five particle-glued bases, where the $x$ and $z$ coordinates are normalized by initial length $L_i$ and initial height $H_i$, respectively. Runout distance decreases monotonically as $R_a$ increases, confirming the effectiveness of using $R_a$ to characterize the base roughness. It is interesting to point out that for Base 2 and Base 3, even though the size and spatial distribution of base particle are different, their runout distances and flow profiles are nearly the same. This observation can be characterized by their similar $R_a$ values in Table 1, which indicates that the two bases provide almost the same frictional resistance to flow body. The tip morphology varies clearly with the base roughness (especially for $x_i/L_i > 1.25$), while the residual height at $x_i/L_i = 0$ only shows minor differences. These basal effects on the deposit morphology can be attributed to the fact that the granular flow only interacts with the base in the flowing region close to the front, as was observed during the column collapse tests. Moreover, when $R_a$ is small, the deposit profile shows more curvature near the tip, while when $R_a$ is large, the profile is roughly triangular.

To further demonstrate the advantage of using $R_a$ for base roughness characterization, we first plot the normalized runout distance against a conventional roughness parameter ($d_b$) in Fig. 8a. Although the runout distance generally increases as $d_b$ decreases, $d_b$ fails to distinguish the influence of the spatial distribution of glued particles (or the spacing between bumps), which may lead to misinterpretation of the basal effect on the runout distance (see red square in Fig. 8a). Moreover, $d_b$ alone cannot explain the same runout distance obtained as $d_b$ decreases (see dashed box in Fig. 8a)

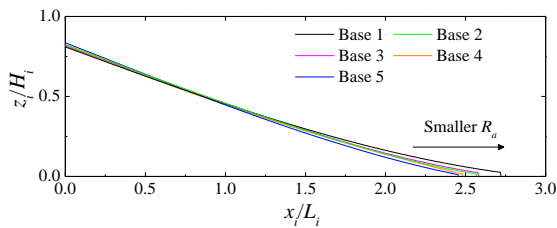

Figure 7. Final deposit profiles for four base types.

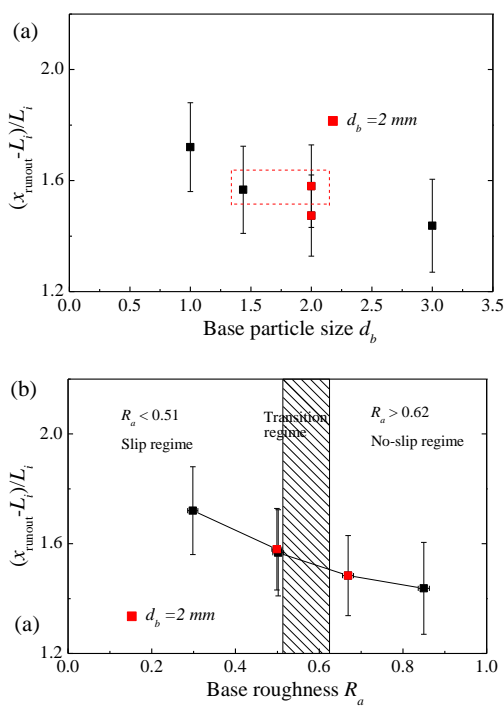

Figure 8. Runout distance vs. different roughness parameters, including (a) the base particle diameter $d_b$ and (b) the base roughness $R_a$ defined in this work.

On the other hand, Fig. 8b presents the normalized runout distance as a function of $R_a$, showing a clear monotonic dependence. Since $R_a$ considers both the base particle size and spacing (or compactness) in a unified manner, it may potentially serve as a roughness indicator for arbitrarily built rough bases using glued particle layers (Jing et al. 2016). It can be observed that the runout distance generally decreases as $R_a$ is increased, but the trend tends to plateau for $R_a = 0.678$ and $0.872$. This roughness effect on the runout distance is reminiscent of a phase transition reported in our DEM simulations of mono-disperse flows (Jing et al. 2016), where the boundary condition can be regarded as no-slip when $R_a > 0.62$, while different degree of slip occurs when $R_a < 0.51$; the slip velocity decreases rapidly in-between these two regimes (see phase transition indicated in Fig. 8b). These finding highlights that the concept of $R_a$, as well as the phase diagram developed in Jing et al. (2016), is not only valid in steady flows, but also in transient flows like the granular column collapse in this work. However, despite the similar transitional behavior that can be characterized using $R_a$, we note an interesting observation that temporary basal slip can occur during the transient column collapse tests even for $R_a > 0.62$. Further investigation, including DEM simulations of granular collapse tests, is needed to fully understand the roughness-induced phase transition in transient granular flows.

## 5 CONCLUSIONS

This work presents the effect of geometric base roughness on the runout distance and deposit morphology in granular column collapse. Based on image processing, robust and accurate measurement of the flow characteristics and base roughness were obtained. A roughness indicator $R_a$ was used to represent a variety of geometric roughness due to different sizes and spatial distributions of base particles. By comparing the cases for different base roughness, it reveals that $R_a$ can effectively reflect the frictional resistance to the flowing particles. When $R_a$ is small, the front of the granular flow can propagate further with less friction at the base. In contrast, when $R_a$ is large, the granular collapse results in a shorter runout distance with a triangular final deposit shape. Moreover, when $R_a$ is nearly the same, the runout distance and deposit profile show little difference. Furthermore, using the base particle size $d_b$ alone fails to quantify the roughness effect due to varying compactness of the base particles, while using $R_a$ produces a monotonic trend for the runout distance results owing to the unified consideration of the size and spacing of bumps (glued particles) at the base.

In future work, a more comprehensive experimental plan will be conducted with various parameters of the granular column collapse tests, such as the column size, aspect ratio, and slope angle, to gain a fuller understanding of the basal effect in granular flows. Also, more sophisticate image processing tools, including the particle image velocimetry, will be adopted to provide detailed flow information and to validate companion DEM simulations. Extraction of particle-level information (e.g., contact force and slip velocity), as well as coarse-grained flow kinematics and stress fields especially near the base, will allow us to develop microscopically informed constitutive relations for the boundary condition of transient granular flows (e.g., basal friction and basal slip laws). The findings are expected to contribute to up-scaled modeling of realistic granular flows in the context of geo-risk assessment and mitigation.

## 6 ACKNOWLEDGEMENTS

This research is supported by Research Grants Council of Hong Kong (General Research Fund No. 17205821).